# The Illusion of Requirements in Software Development[1]

Paul Ralph

Lancaster University

paul@paulralph.name | http://paulralph.name

It is widely accepted that understanding system requirements is important for software development project success [1,2]. Put another way, it is widely acknowledged that failing to understand requirements is related to project failure [3,4]. The idea that software artifacts generally have a set of discoverable, documentable requirements is entrenched in industry standards [5], development processes [2] and educational curricula [6,7]. More broadly, requirements are a fundamental component of the Rational Model of Design [8-10], the dominant view of how practitioners approach developing software and information systems. However, utilizing good requirements practices may not be a necessary or sufficient condition for project success [11,12].

The assumption that software projects have discoverable, documentable requirements has motivated diverse literature on requirements engineering (RE), the process of identifying, analyzing, modeling, verifying and managing requirements. Major contributions include RE approaches (e.g., goal-oriented RE [13], user-centered RE [14]) and requirement types (e.g., non-functional requirements [15], early requirements [16]).

However, at least three software development projects that I have observed or participated in have produced not a single, meaningful requirement. Although they produced statements labelled as requirements, closer inspection revealed that they were something else – goals, design decisions, to-do- or wish-list items. Therefore, the purpose of this viewpoint is to explore the possibility of software projects with few or illusory requirements.

## 1 Some Explicit Assumptions

Different authors have defined "requirement" in different ways, alternatively as "a structural or behavioral property that a design object must possess" [17, p. 108], "a statement that identifies a capability or function that is needed by a system in order to satisfy its customer's needs" [18, p. 205] and "a property which must be exhibited in order to solve some problem in the real world" [19, sec. 1.1].

Requirements are often differentiated from goals and design decisions. "A goal is an objective the system under consideration should achieve" [13, p. 2]. Where "goals describe the desired impacts of a design object on its environment" [17, p. 110], "requirements are usually understood as stating what a system is supposed to do, as opposed to how it should do it" [20, p. 226]. Similarly, Roman argued that "requirements specifications state the desired functional and performance characteristics of some component independent of any actual realization" [21, p. 14].

---

Likewise, IEEE Standard 830-1998 states that "a requirement specifies an externally visible function or attribute of a system [while] a design describes a particular subcomponent of a system and/or its interfaces with other subcomponents" [5, p. 9]. Understanding the relationships between goals, requirements and design decisions is the essence of the requirements traceability problem [cf. 22,23].

These varying definitions clarify the need for more explicit ontological assumptions underlying RE. For the purposes of this paper, I assume that a software development project produces an artifact called the *design object* having properties called *features*. Project participants or stakeholders may assign to the design object one or more goals – optative statements describing a change in the environment that the design object is desired to produce. A *requirement* is a feature of a design object that is necessary to achieve a goal. (While this is an unconventional way of defining *requirement*, it significantly simplifies explaining the two challenges advanced below, which do not hinge on this particular definition.) For example, suppose a development team is designing a website (the design object) to sell cameras online (the goal). The website will have many features (e.g., having the store name in a bold font) that do not substantially contribute to achieving its goal. However, other features (e.g., shopping cart) may be necessary conditions for achieving the goal – these necessary features are requirements. More precisely, given a goal $g$, a set of requirements $R_g$, may be defined as the set of all features necessary for a design object to achieve $g$. For the purposes of this paper I do not distinguish between early- and late-stage requirements [16], hard / functional and soft / non-functional requirements [15] or requirements and constraints [17].

## 2 Ontological and Epistemological Challenges

Suppose exactly two design objects $D_1$ and $D_2$ will achieve a goal $g$. Features of both design objects are requirements as there is no way to achieve $g$ without them (Table 1). Properties of $D_1$ but not $D_2$ are not requirements as $g$ may be achieved without them (building $D_2$); properties of $D_2$ but not $D_1$ are not requirements as g may be achieved without them (building $D_1$). Properties of neither object are irrelevant. More generally, suppose $g$ may be achieved by $n$ design objects ($D_1, D_2, ... D_n$) each with features ($F_1, F_2, ... F_n$). The set of requirements $R_g$ may then be defined as the intersection of the properties ($R_g = F_1 \cap F_2 \cap ... \cap F_n$).

| Table 1. Relationship between properties and requirements | | |
|---|---|---|
| | Property of $D_1$ | Not a Property of $D_1$ |
| Property of $D_2$ | requirement | feature |
| Not Property of $D_2$ | feature | irrelevant |

Characterizing requirements as such reveals two assumptions. First, the existence of requirements entails overlap among design object features that may achieve a goal ($R_g = F_1 \cap F_2 \cap ... \cap F_n \neq \varnothing$). In contrast, if the properties do not overlap ($R_g = F_1 \cap F_2 \cap ... \cap F_n = \varnothing$), there are no requirements (an ontological problem; Fig. 1). Second, stating requirements assumes that all relevant design objects (or classes thereof) have been identified. If another solution, which will achieve *g* without any properties in common with known solutions, may exist, no properties of the existing solution sets are requirements (an epistemological problem). For example, suppose a company sets out to protect their network against malware, writing a requirements specification for a sophisticated antivirus system. During review, someone asks "Why don't we just switch to Mac OS or Linux instead as they have fewer problems with viruses?". This reveals ostensible 'requirements' as merely features of one solution.

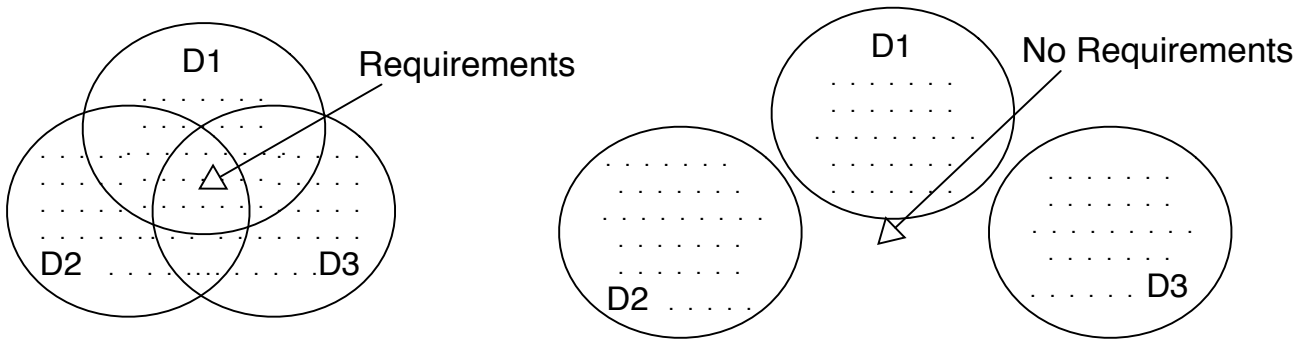

**Figure 1.** Relationship between features (dots) and requirements.

Note: features common to all design alternatives are requirements (left); where no features overlap, no requirements exist (right).

In summary, if the solution space is unknowable or lacks overlap, no requirements can be stated. This is not the same as acknowledging ambiguous, conflicting or incomplete requirements. Either lack of overlap or epistemic uncertainty produce a no-requirements scenario.

## 3. Limitations and Proceeding without Requirements

The ontological and epistemological challenges to the requirements concept described above are obviously somewhat simplified. First, the discussion takes a counterfactual approach to causality [24] when a probabilistic approach would be more appropriate (features may increase success probability rather than being "necessary conditions"). Second, the definition of requirement excludes other non-critical desiderata (wants, preferences), which may still be important. Third, identifying a needed feature may not always necessitate understanding the full scope of the design space – a highly credible informant's comment that 'the board will never approve this if it won't work on an iPhone' may be sufficient to justify stating a requirement.

However, the illustration illuminates two fundamental problems. First, we can conceive of situations (e.g., knowledge worker burnout) where two completely different approaches (e.g., better tool support or hire more employees) may achieve the same goal (e.g., decrease work hours) – in such cases of low solution overlap, few if any requirements can be stated. Second, without fully exploring the design space we cannot be sure whether there exists another approach, which would achieve the goal without any features of known approaches.

This leaves the intellectual enterprise of RE research with two possibilities. One is that we should expect many software development projects to have few if any legitimate requirements, rendering many requirements engineering approaches ineffective or inappropriate in these contexts. This raises fundamental questions concerning how to adapt existing RE approaches to no-requirements scenarios, or how to proceed without requirements more generally.

Possibility two is that many existing RE approaches operationalize the requirements construct more generally – as desiderata, which may or may not be strictly necessary for success. This is problematic due to the strong denotation and connotation of the term. The word "requirement" denotes a thing that is compulsory. Listing requirements connotes certainty and unambiguousness. For example, when an analyst states that 'cross-platform compatibility' is a requirement, novice developers and stakeholders unfamiliar with the challenges of RE are unlikely to interpret this as 'the analyst hypothesizes that cross-platform compatibility will increase the probability that the system will achieve its objectives but we will not know for sure until the system is built, if ever' or 'all of the plausible design candidates generated so far include cross-platform compatibility, but we have not fully explored the design space'. A wide variety of cognitive phenomena including anchoring bias [25], fixation [26] and confirmation bias [27] suggest that misrepresenting an incidental feature as a requirement will reduce exploration of the design space, curtailing innovation [28].

The argument of this paper may be challenged on at least two grounds. First, one may draw a distinction between mandatory and optional requirements. In response I reiterate the psychological effect of mislabeling features as "requirements" (above), and suggest that the "optional requirement" label will likely increase confusion. Second, one may argue for conditional requirements, wherein some features become requirements conditional on other features, e.g., 'given that the design artifact is a website, it must be HTML5-compliant'. However, this creates a reductionist spiral where virtually all design decisions may be recast as conditional requirements, undermining the distinction between requirements and design decisions and further confusing developers and curtailing innovation.

In conclusion, this paper presents two novel challenges to RE. The ontological challenge posits that where many plausible approaches to achieving a goal are evident, there may be insufficient overlap between approaches to form requirements. The epistemological challenge posits that while all plausible approaches may have sufficient overlap to state requirements, one cannot know that unless all approaches are identified and one is somehow sure that none have been missed. I have formulated these challenges to stimulate debate on fundamental properties and assumptions of requirements in theory and practice. They raise important questions about possible requirement-sparse environments and the implications of goals, features, conjectures and design decisions mislabeled as requirements.

# Bibliography

1. Boehm B (1991) Software Risk Management: Principles and Practices. IEEE Software 8 (1):32-41
2. Jacobson I, Booch G, Rumbaugh J (1999) The Unified Software Development Process. Addison-Wesley Longman Publishing Co., Inc., Boston, MA, USA
3. Ewusi-Mensah K (2003) Software Development Failures. MIT Press, Cambridge, MA, USA
4. Standish Group (2009) CHAOS Summary 2009. Boston, MA, USA http://www.standishgroup.com/newsroom/chaos_2009.php
5. IEEE (1998) IEEE Standard 830-1998: Recommended Practice for Software Requirements Specifications.
6. Topi H, Valacich JS, Wright RT, Kaiser KM, Nunamaker JF, Sipior JC, Vreede GJd (2010) IS 2010: Curriculum Guidelines for Undergraduate Degree Programs in Information Systems. Communications of Association for Information Systems 26
7. Joint Task Force on Computing Curricula (2004) Software Engineering 2004: Curriculum Guidelines for Undergraduate Degree Programs in Software Engineering. Díaz-Herrera JL, Hilburn TB (eds) http://sites.computer.org/ccse/SE2004Volume.pdf
8. Simon HA (1996) The Sciences of the Artificial. 3rd edn. MIT Press, Cambridge, MA, USA
9. Brooks FP (2010) The Design of Design: Essays from a Computer Scientist. Addison-Wesley Professional
10. Ralph P (2011) Introducing an Empirical Model of Design. Paper presented at the The 6th Mediterranean Conference on Information Systems, Limassol, Cyprus, Sept. 3-5
11. Davis AM, Zowghi D (2006) Good Requirements Practices are Neither Necessary nor Sufficient. Requirements Engineering 11 (1):1-3
12. Shenhar AJ, Dvir D, Levy O, Maltz AC (2001) Project Success: A Multidimensional Strategic Concept. Long Range Planning 34 (6):699-725
13. van Lamsweerde (2001) A Goal-oriented Requirements Engineering: a Guided Tour. In: Proceedings of the Fifth IEEE International Symposium on Requirements Engineering, Aug, pp 249-262
14. Sutcliffe A, Thew S, Jarvis P (2011) Experience with User-Centred Requirements Engineering. Requirements Engineering 16 (4):267-280
15. Chung L, Nixon BA, Yu E (2000) Non-Functional Requirements in Software Engineering. Kluwer International Series in Software Engineering, vol 5. Springer
16. Fuxman A, Liu L, Mylopoulos J, Pistore M, Roveri M, Traverso P (2004) Specifying and Analyzing Early Requirements in Tropos. Requirements Engineering 9 (2):132-150
17. Ralph P, Wand Y (2009) A Proposal for a Formal Definition of the Design Concept. In: Lyytinen K, Loucopoulos P, Mylopoulos J, Robinson W (eds) Design Requirements Engineering: A Ten-Year Perspective. Lecture Notes on Business Information Processing, vol 14. Springer-Verlag, pp 103-136
18. Bahill AT, Dean FF (2009) Discovering System Requirements. In: Sage AP, Rouse WB (eds) Handbook of Systems Engineering and Management. 2nd edn. John Wiley & Sons, pp 205-266
19. Bourque P, Dupuis R (eds) (2004) Guide to the Software Engineering Body of Knowledge (SWEBOK). IEEE Computer Society Press
20. Yu E (1997) Towards Modelling and Reasoning Support for Early-Phase Requirements Engineering. In: Proceedings of the Third IEEE International Symposium on Requirements Engineering, pp 226-235
21. Roman G-C (1985) A Taxonomy of Current Issues in Requirements Engineering. Computer 18 (4):14-23
22. Gotel O, Finkelstein A An analysis of the requirements traceability problem. In: First International Conferance on Requirements Engineering, Colorado Springs, CO, USA, 1994. IEEE Computer Society Press, pp 94-101
23. Ramesh B, Jarke M (2001) Toward Reference Models for Requirements Traceability. IEEE Trans Softw Eng 27 (1): 58-93
24. Gregor S (2006) The Nature of Theory in Information Systems. MIS Quarterly 30 (3):611-642
25. Parsons J, Saunders C (2004) Cognitive Heuristics in Software Engineering: Applying and Extending Anchoring and Adjustment to Artifact Reuse. IEEE Transactions on Software Engineering 30:873-888
26. Jansson DG, Smith SM (1991) Design Fixation. Design Studies 12 (1):3-11.
27. Oswald ME, Grosjean S (2004) Confirmation Bias. In: Pohl RF (ed) Cognitive Illusions: A Handbook on Fallacies and Biases in Thinking, Judgement and Memory. Psychology Press, Hove, UK, pp 79-96
28. Ralph P (2011) Toward a Theory of Debiasing Software Development. In: Wrycza S (ed) Research in Systems Analysis and Design: Models and Methods: 4th SIGSAND/PLAIS EuroSymposium 2011. LNBIP, vol 93. Springer, Gdansk, Poland, pp 92-105